\documentstyle[aps,prl,preprint]{revtex}
\draft
\begin{document}
\title{
Multiscale Reference Function  Analysis of the
${\cal P}{\cal T}$ Symmetry Breaking Solutions for the  $P^2+iX^3+i\alpha X$
 Hamiltonian}
\author{C. R. Handy$^1$, D. Khan$^1$, Xiao-Qian Wang$^1$,
 and C. J. Tymczak$^2$  }
\address{$^1$Department of Physics \& Center for Theoretical Studies of 
Physical Systems, Clark Atlanta University, 
Atlanta, Georgia 30314}
\address{$^2$ Los Alamos National Laboratory T-14, Los Alamos, New Mexico 87545}
\date{Received \today}
\maketitle
\begin{abstract}
The recent work of Delabaere and Trinh (2000 J. Phys. A 33 8771) discovered the
existence of ${\cal P}{\cal T}$-symmetry breaking, complex energy, $L^2$
 solutions
for the one dimensional  Hamiltonian, $P^2+iX^3+i\alpha X$, in the 
asymptotic limit, $\alpha \rightarrow -\infty$. Their asymptotic
analysis produced questionable results for 
  moderate values of  
$\alpha$. We can easily confirm the existence of
${\cal P}{\cal T}$-symmetry breaking solutions, by explicitly 
computing the low lying states,
for $|\alpha| < O (10)$. Our analysis makes  use of
the  Multiscale Reference Function (MRF) approach,
developed by Tymczak et al (1998 Phys. Rev. Lett. 80 3678; 1998 Phys. Rev. A 58, 2708). 
The MRF results can be validated by comparing them
with the
 converging eigenenergy bounds generated 
 through  the Eigenvalue Moment Method, as recently argued by
Handy (2001a,b).
 Given the reliability of the MRF analysis,
 its fast numerical implementation,  high 
accuracy, and theoretical simplicity, the present formalism defines
an effective and efficient  procedure
 for analyzing many related problems that 
have appeared in the recent literature.

\end{abstract}
\vfil\break
\section{Introduction}

There has been much interest, recently, in understanding the symmetry breaking
mechanism for ${\cal P}{\cal T}$ invariant Hamiltonians of the type
$P^2 + \sum_{j=0}^JC_J (iX)^j$. The mathematical 
interest in these systems
originated from a conjecture by D. Bessis, and WKB analysis
 confirmation by Bender and
Boettcher (1998), that the class of potentials of the
form $V(x) = (iX)^n$ only allow for ${\cal P}{\cal T}$-invariant
solutions, and thus can only have  real discrete spectra.
  The recent literature 
testifies to the great interest in these problems, as can be found in the
cited references by Bender et al (2001), and Mezincescu (2000, 2001). 

  An important  work establishing that ${\cal P}{\cal T}$-symmetry breaking
systems do exist 
 was the recent study by Delabaere and
 Trinh (2000) which used asymptotic methods to analyze the Hamiltonian
$H_\alpha \equiv P^2 + iX^3 + i\alpha X$. Their analysis showed the existence
of symmetry breaking solutions for large $\alpha$ values; however, for moderate
values, their results were near the limits of their analytical validity.
Despite this, they made certain questionable
 predictions for moderate $\alpha$ values.

Our immediate objective is to check the validity of the 
 Delabaere and
Trinh study, 
by explicitly computing the low-lying complex (and real) eigenenergies for 
moderate
$\alpha$ values. We do this in two ways. The first makes use of the very
efficient Multiscale Reference Function (MRF)
 formalism of Tymczak et al (1998a,b).
The results of this eigenenergy estimation 
analysis are then confirmed through application of a
recently developed eigenenergy bounding theory proposed by Handy (2001a,b), 
which
can generate converging  bounds to the complex eigenenergies. This
bounding approach, referred to as the Eigenvalue Moment Method (EMM), is 
exact, although numerically slower in its implementation than the
MRF procedure.

As a footnote to the above, we emphasize that 
the EMM theory generates an infinite hierarchy of closed,
finite dimensional,
algebraic, eigenenergy constraints. These are then solved, numerically.
In this regard, the EMM procedure is very different from other numerical
schemes, such as numerical integration, which are intrinsically of an
approximating nature, and cannot provide any fundamental
 theoretical insight into
the underlying physical processes. The algebraic constraints generated
through the EMM formalism can provide such insight; although such analysis
have not been attempted, heretofore. However, for the immediate purposes
of this study, we solely defer to EMM in order to check the validity of the 
MRF results.

In this work, 
we provide the essentials of the MRF theoretical structure,
as applied to the $H_\alpha$ Hamiltonian. The EMM theory is not discussed.
Only the numerical bounds are quoted in the Tables.
This work   validates the relevancy of MRF theory in the
computation of complex eigenenergies, for the class of problems referenced
above.

Both the MRF and EMM methods
 are dependent on a moments' representation for the given system. This in turn
is
 readily realizable for any (multidimensional) rational fraction potential.

Any moment based analysis is inherently multiscale in nature. That is, as the
number of  moments used increases, one is probing the system at successively
smaller scales. 
Consistent with this, the MRF basis 
representation, particularly within configuration space,
 has important
ties with (complex) 
turning point quantization (Handy et al (2000)), and wavelet analysis
(Handy and Brooks (2001)).

We outline the basic MRF theory and its implementation, in the next section.
 The last section contains a detailed enumeration, and illustration, of the MRF results, which make 
precise the qualitative spectral structure conjectured by Delabaere and Trinh.

\vfil\break
\section{ The MRF Representation}

\subsection {The Moment Equation}
The starting point for the MRF analysis is the transformation of the
Schrodinger equation into the Fourier representation, assuming that
one is working with the physical, $L^2$, solutions. Thus, for the
configuration space Schrodinger equation studied by Delabaere and
Trinh (2000)

\begin{equation}
-\partial_x^2\Psi(x) + (ix^3 + i\alpha x) \Psi(x) = E \Psi(x),
\end{equation}
its Fourier transform counterpart is (i.e. $x \rightarrow i\partial_k$,
$\partial_x \rightarrow ik$)

\begin{equation}
k^2 {\hat{\Psi}}(k) + (\partial_k^3 - \alpha \partial_k){\hat{\Psi}}(k) = E
{\hat{\Psi}}(k),
\end{equation}
where
\begin{equation}
{\hat{\Psi}}(k) = {1\over {\sqrt{2\pi}}} \int_{-\infty}^{+\infty}dx \ e^{-ikx}
\Psi(x).
\end{equation}

It is important to note that, for the physical solutions, a simple application of WKB analysis (Bender and Boettcher (1998)) tells us that the asymptotic 
behavior of the configuration space representation yields an entire Fourier
transform. Because of this, the $k$-power series expansion is absolutely convergent, and defined in terms of the power moments:

\begin{equation}
{\hat{\Psi}}(k) = {1\over {\sqrt{2\pi}}}\sum_{p=0}^\infty (-ik)^p{{\mu_p}\over {p!}},
\end{equation}
where
\begin{equation}
\mu_p \equiv \int_{-\infty}^{+\infty} dx \ x^p \Psi(x),
\end{equation}
define the Hamburger power moments.

We can generate the recursion relation for the $\mu_p$'s from the standard 
power series expansion methods for linear differential equations (Bender and
Orszag (1978)). Alternatively, we can apply $\int_{-\infty}^{+\infty} dx \ x^p$
to both sides of Eq.(1), combined with integration by parts, and obtain the
necessary {\it Moment Equation}:

\begin{equation}
\mu_{p+3} = -\alpha \mu_{p+1} -i E \mu_{p} - i p(p-1)\mu_{p-2},
\end{equation}
for $p \geq 0$. This corresponds to a homogeneous, linear,
 finite difference equation, of effective
order $1+m_s = 3$, since specification of the independent moments 
$\{\mu_0,\mu_1,\mu_2\}$, plus the (complex) energy parameter, $E$,
generates all of the remaining moments. The independent moments are
referred to as the {\it missing moments}.

The linear dependence of the moments, on the missing moments, can be
expressed through the relation

\begin{equation}
\mu_p = \sum_{\ell = 0}^{m_s} M_{p,\ell}(E) \mu_\ell,
\end{equation}
where the energy dependent coefficients satisfy the moment equation, with
respect to the $p$-index, as well as the initial conditions
$M_{\ell_1,\ell_2} = \delta_{\ell_1,\ell_2}$, for
$0 \leq \ell_1,\ell_2 \leq m_s$.

\subsection{ Defining an Analytic Basis in the Fourier Space}

The physical solutions in the Fourier representation must also be $L^2$.
One would like to find an appropriate basis into which to transform the
Fourier power series expansion:

\begin{equation}
{1\over {\sqrt{2\pi}}}\sum_{p=0}^\infty (-ik)^p{{\mu_p}\over {p!}}
 = {1\over {\sqrt{2\pi}}} \sum_{j=0}^\infty a_j{\cal B}_j(k).
\end{equation}
The easiest choice, leading to a rapid, analytic generation of the 
$a_j$ coefficients, is to take 

\begin{equation}
{\cal B}_j(k) = (-ik)^j {\hat R}(k),
\end{equation}
where ${\hat R}$
 is some arbitrary ``reference" function yielding a complete (if not
orthogonal) basis. Thus, if ${1\over{{\hat R}(k)}}$ is analytic, then one
can generate the $a_j$'s by expanding ${{{\hat{\Psi}}(k)}\over{{\hat R}(k)}}$.
In particular, we can take ${\hat R}(k) = e^{-\beta k^2}$, where $\beta > 0$,
and otherwise arbitrary, yielding
\begin{equation}
A(k) \equiv \sum_{j=0}^\infty a_j (-ik)^j = e^{\beta k^2}\sum_{p = 0}^\infty 
(-ik)^p{{\mu_p}\over {p!}}.
\end{equation}

It is then clear that the $a_j$'s will become linear in the missing moments,
$\{\mu_\ell|0 \leq \ell \leq m_s\}$. Specifically,

\begin{equation}
a_j(E;\mu_0,\mu_1,\mu_2) = \sum_{p+2q = j} {{(-\beta)^q \mu_p}\over {q! p!}},
\end{equation}
or
\begin{equation}
a_j(E;\mu_0,\mu_1,\mu_2) = \sum_{\ell = 0}^{m_s = 2}  
\Big(\sum_{p+2q = j} {{(-\beta)^q M_{p,\ell}(E)}\over {q! p!}}\Big )
\ \times \ \mu_\ell,
\end{equation}
for $j \geq 0$. Clearly, the maximum (Hamburger) moment order generated, $P_{max}$,
determines the maximum order of $a_j$'s generated, $0 \leq j \leq P_{max}$.

\subsection{ The MRF Quantization Prescription}

It has been argued by Tymczak et al (1998a,b) that the convergent zeroes
of the coefficient functions

\begin{equation}
a_j(E^{(j)}_n) = 0,
\end{equation}
converge to the exact discrete state energies:

\begin{equation}
\lim_{j\rightarrow \infty} E^{(j)}_n = E^{physical}_n.
\end{equation}

The above root equation must be adapted to the, $1+m_s$, linear, missing moment 
structure of the Hamiltonian in question. Thus, to any expansion order $J$, we
impose that the last $1+m_s$, $a_j$-coefficients be zero (i.e. $a_{J-\ell} = 0$,
for $0 \leq \ell \leq m_s$). This results in a
$1+m_s = 3$ dimensional, determinantal equation for the energy:

\begin{equation}
\Delta_J(E) = 0,
\end{equation}
where
\begin{equation}
\Delta_J(E) \equiv Det\pmatrix{ 
{\cal M}_{0,0}^{(J)}(E,\beta), {\cal M}_{0,1}^{(J)}(E,\beta),{\cal M}_{0,2}^{(J)}(E,\beta) \cr
{\cal M}_{1,0}^{(J)}(E,\beta), {\cal M}_{1,1}^{(J)}(E,\beta),{\cal M}_{1,2}^{(J)
}(E,\beta) \cr
{\cal M}_{2,0}^{(J)}(E,\beta), {\cal M}_{2,1}^{(J)}(E,\beta),{\cal M}_{2,2}^{(J)
}(E,\beta) \cr},
\end{equation}
and
\begin{equation}
 {\cal M}_{\ell_1,\ell_2}^{(J)}(E,\beta) \equiv 
\sum_{p+2q = J-\ell_1} {{(-\beta)^q M_{p,\ell_2}(E)}\over {q! p!}},
\end{equation}
for $0 \leq \ell_1,\ell_2 \leq m_s = 2$. In the Tables, the $P_{max}$
parameter corresponds to $J \equiv P_{max}$.

\vfil\break
\section {Numerical Implementation of MRF}

The MRF analysis is implemented for $\beta = .5$. This value of the arbitrary
parameter generates the fastest converging results.

In Figure I we plot the ${\cal P}{\cal T}$-symmetry breaking solutions ($Im(E) \neq 0$), and the ${\cal P}{\cal T}$-symmetry invariant solutions
 ($Im(E) = 0$), for moderate 
$\alpha$ values at the limits of  Delabaere and Trinh's asymptotic
 analysis.
 There
are four branches depicted.
In Tables I-IV we specify some of the points 
plotted, and in addition, compare the MRF eigenenergy estimates with the
EMM bounds.  The MRF results in these tables generally correspond to $J \leq 50$. The EMM results were calculated to the same
 moment order, $P_{max} \leq 50$, for the complex energies. In those cases where the MRF method predicts
a purely real eigenenergy, we used the faster EMM formalism corresponding
to explicitly assuming that the underlying solutions are ${\cal P}{\cal T}$
invariant (Handy (2001a)). For this case, the underlying problem is of
Stieltjes character, and a Stieltjes moment order  of $P_{max}^{(S)} = 30$,
 corresponds to
a Hamburger moment order of 60 (i.e. $P_{max}^{(S)} = {{P_{max}}\over 2}$).
 For this reason,  in
some cases the MRF estimates (computed at $J \leq 50$)
 lie outside of the EMM bounds. However,
for some of the EMM bounds, numerical instability concerns required that
$P_{max}^{(S)} < 30$ (these were computed on a standard, double precision,
IBM platform). Such cases are relatively few in number, and are explicitly
identified. By working at a larger precision order, the corresponding bounds
can be improved.

It will also be noted that for $\alpha = 0$, we quote the EMM bounds generated
at higher Stieltjes moment order ($P_{max}^{(S)} \leq 60$), given in the
work by Handy (2001a). These bounds are very consistent with the corresponding
entries in Tables V-VI.

In Tables V and VI we quote the MRF results, for $J = 100$ (using CRAY-double
precision). Only the stable digits are given (that is, Tables V and VI correspond
to the (empirically determined) stable digits within the MRF generated sequence,
for $0 \leq J \leq 100$).

  The results are consistent, that is, the MRF results in Tables I-IV, for the
most part, lie within
the bounds. This is definitely the case for Tables V - VI. 
Note that in the Tables we quote the imaginary parts of the energy
as $\pm$, this is because we cannot tell which ${\cal P}{\cal T}$-invariant 
branch continues
into the ${\cal P}{\cal T}$-breaking  branch. For ${\cal P}{\cal T}$ 
invariant Hamiltonians, complex energies come in conjugate pairs. Thus, if
$E$ is a solution, so too is $E^*$.

In Figures II - III we narrow in on the smaller of
 the two critical $\alpha$ values,
$\alpha_{cr_{1,2}}$, as shown in Fig. I.
At these critical points, the energy goes from being real (${\cal P}{\cal T}$ - 
invariant solutions) to complex (${\cal P}{\cal T}$ - breaking solutions). They
are:
\begin{equation}
\alpha_{cr_1} = -2.611809356,
\end{equation}
corresponding to $E_{cr_1} = 1.28277353562$; and
\begin{equation}
\alpha_{cr_2} = -5.375879629,
\end{equation}
corresponding to $E_{cr_2} = 4.181388093.$

\begin{table}
\caption {$E_1$-Branch for the Discrete States of $P^2 + iX^3 + i\alpha X$}
\begin{center}
\begin{tabular}{cccccl}
\multicolumn{1}{c}{$\alpha$}& \multicolumn{1}{c}{$E_1^{(MRF)}(\alpha)$} & \multicolumn{1}{c}{$EMM Bounds: E_R^{(L)} < E_R  < E_R^{(U)}$} \& {$E_I^{(L)} < E_I< E_I^{(U)}$}\\ \hline

 -5.0 & (    1.3433409,    $\pm$ 2.9073602) & $1.343311^1 < E_R <1.343354^1$,
 $2.9073^1 < E_I < 2.9075^1$         \\
 -4.5 & (    1.2992519,    $\pm$ 2.3124924) & $1.299242 <E_R < 1.299252$,        $2.3124 < E_I < 2.3126$ \\
 -4.0 & (    1.2486637,    $\pm$ 1.7617076) & $1.248637 < E_R < 1.248666$,
$1.761688 < E_I < 1.761742$         \\
 -3.5 & (    1.2124399,    $\pm$ 1.2609114) & $1.212421 < E_R < 1.212448$,
$1.26088 < E_I < 1.26094$        \\
 -3.0 & (    1.2258438,    $\pm$ 0.7600296) & $1.225837 < E_R < 1.225864$,
$.76000 < E_I < .76004$         \\
 -2.5 & (     .9280136, 0 ) & $  0.92799980^2  < E < 0.92800101^2 $\\
 -2.0 & (     .6209137, 0 ) & $  0.62091347    < E < 0.62091386 $        \\
 -1.5 & (     .5964936, 0 ) & $  0.59649326    < E < 0.59649351 $        \\
 -1.0 & (     .6999615, 0 ) & $  0.69995977    < E < 0.69995978 $  \\
  -.5 & (     .8926699, 0 ) & $  0.89266827    < E < 0.89266849  $        \\
   .0 & (    1.1562673, 0 ) & $  1.15626695   < E <  1.15626718 $        \\
      &                  & $1.15626707198811324^3 < E < 1.15626707198811335^3$\\
   .5 & (    1.4798519, 0 ) & $  1.47985179    < E < 1.47985206 $        \\
  1.0 & (    1.8561128, 0 ) & $  1.85611065    < E < 1.85611108 $       \\
  1.5 & (    2.2797563, 0 ) &  $ 2.27975185    < E < 2.27975232 $       \\
  2.0 & (    2.7467434, 0 ) &  $ 2.74673952    < E < 2.74674023 $       \\
  2.5 & (    3.2538767, 0 ) &  $ 3.25387596 < E <    3.25387723 $       \\
  3.0 & (    3.7985559, 0 ) &  $ 3.79855387 < E <    3.79855395 $ \\
  3.5 & (    4.3786140, 0 ) &  $ 4.37859645  < E <   4.37859736 $     \\
  4.0 & (    4.9921974, 0 ) &  $ 4.99215436 < E <    4.99215504 $ \\
  4.5 & (    5.6376822, 0 ) &  $ 5.63763149 < E <    5.63763200 $     \\
  5.0 & (    6.3136428, 0 ) &  $ 6.31359739 < E <    6.31360665 $       

\end{tabular}
\end{center}
\noindent{$^1$EMM Analysis of Handy (2001b) \\ 
$^2$EMM ($P_{max}^{(S)} = 30$) Analysis of Handy (2001a) \\
$^3$EMM ($P_{max}^{(S)} = 60$) Analysis of Handy (2001a) }
\end{table}

\begin{table}
\caption {$E_2$-Branch for the Discrete States of $P^2 + iX^3 + i\alpha X$}
\begin{center}
\begin{tabular}{cccccl}
\multicolumn{1}{c}{$\alpha$}& \multicolumn{1}{c}{$E_2^{(MRF)}(\alpha)$} & \multicolumn{1}{c}{$EMM Bounds: E_R^{(L)} < E_R  < E_R^{(U)}$} \& {$E_I^{(L)} < E_I< E_I^{(U)}$}\\ \hline

 -5.0 & (    1.3433409,    $\pm$ 2.9073602) & $1.343311^1 < E_R <1.343354^1$,
 $2.9073^1 < E_I < 2.9075^1$         \\ 
 -4.5 & (    1.2992519,    $\pm$ 2.3124924) & $1.299243 <E_R < 1.299252$,
 $2.3124 < E_I < 2.3126$        \\
 -4.0 & (    1.2486637,    $\pm$ 1.7617076) &  $1.248637 < E_R < 1.248666$,
$1.761688 < E_I < 1.761742$         \\
 -3.5 & (    1.2124399,    $\pm$ 1.2609114) &  $1.212421 < E_R < 1.212448$,
$1.26088 < E_I < 1.26094$       \\
 -3.0 & (    1.2258438,    $\pm$ 0.7600296) &  $1.225837 < E_R < 1.225864$,
$.76000 < E_I < .76004$       \\
 -2.5 & (    1.6859358, 0 ) &  $1.68597765^2 < E <    1.68598087^2$    \\
 -2.0 & (    2.2922626, 0 ) &  $2.29229055 < E <    2.29229333 $   \\
 -1.5 & (    2.7425268, 0 ) &  $2.74252667 < E <    2.74253034  $     \\
 -1.0 & (    3.1797220, 0 ) &  $3.17971312 < E <    3.17971750  $    \\
  -.5 & (    3.6320373, 0 ) &  $3.63207237 < E <    3.63207767 $        \\
   .0 & (    4.1091279, 0 ) &  $4.10922704 < E <    4.10923558 $    \\
      &                     & $4.109228752806^3 < E < 4.109228752812^3$ \\
   .5 & (    4.6147402, 0 ) &  $4.61483391 < E <    4.61484633 $  \\
  1.0 & (    5.1501688, 0 ) &  $5.15016059 < E <    5.15017640 $     \\
  1.5 & (    5.7154576, 0 ) &  $5.71538438 < E <    5.71541649 $       \\
  2.0 & (    6.3100192, 0 ) &  $6.31020527 < E <    6.31025282 $ \\
  2.5 & (    6.9332376, 0 ) &  $6.93405453 < E <    6.93412954 $\\
  3.0 & (    7.5850094, 0 ) &  $7.58627841 < E <    7.58638028 $\\
  3.5 & (    8.2656580, 0 ) &  $8.26613065 < E <    8.26633647 $\\
  4.0 & (    8.9745543, 0 ) &  $8.97272640^4 < E <    8.97332836^4 $   \\
  4.5 & (    9.7077326, 0 ) &  $9.70210565^4 < E <    9.70789436^4 $  \\
  5.0 & (   10.4575130, 0 ) & $10.45227656^4 < E <   10.48006875^4 $        

\end{tabular}
\end{center}
\noindent{$^1$EMM Analysis of Handy (2001b) \\ 
$^2$EMM ($P_{max}^{(S)} = 30$) Analysis of Handy (2001a) \\
$^3$EMM ($P_{max}^{(S)} = 50$)  Analysis of Handy (2001a) \\
$^4$EMM ($P_{max}^{(S)} < 30$) Analysis of Handy (2001a)}
\end{table}

\begin{table}
\caption {$E_3$-Branch for the Discrete States of $P^2 + iX^3 + i\alpha X$}
\begin{center}
\begin{tabular}{cccccl}
\multicolumn{1}{c}{$\alpha$}& \multicolumn{1}{c}{$E_3^{(MRF)}(\alpha)$} & \multicolumn{1}{c}{$EMM Bounds: E_R^{(L)} < E_R  < E_R^{(U)}$} \& {$E_I^{(L)} < E_I< E_I^{(U)}$}\\ \hline

 -7.5 & (    4.4103093,    $\pm$ 3.2664292) & $4.408662^1 < E_R < 4.411426^1$,
$3.2645 < E_I < 3.2700$        \\
 -7.0 & (    4.3120349,    $\pm$ 2.5634751) & $4.310530 < E_R < 4.313107$, 
$2.5618 < E_I < 2.5674$        \\
 -6.5 & (    4.2056380,    $\pm$ 1.9039658) & $4.204520 < E_R < 4.206360$,
$1.9026 < E_I < 1.9074$         \\
 -6.0 & (    4.1343003,    $\pm$ 1.2850914) & $4.133864 < E_R < 4.134534$, $1.284 < \pm E_I < 1.286$         \\
 -5.5 & (    4.1575135,    $\pm$ 0.5313291) &  $4.155648 < E_R < 4.158235$,
$.5300 < E_I < .5370$        \\
 -5.0 & (    3.4314015, 0 ) & $3.43136741^2 < E <  3.43139540^2 $       \\
 -4.5 & (    3.3232446, 0 ) & $3.32323635   < E <  3.32325018 $\\
 -4.0 & (    3.5087615, 0 ) & $3.50876099   < E <  3.50877544 $      \\
 -3.5 & (    3.8776981, 0 ) & $3.87768485   < E <  3.87770178 $     \\
 -3.0 & (    4.3334536, 0 ) & $4.33342275   < E <  4.33344654 $   \\
 -2.5 & (    4.8229806, 0 ) & $4.82294849   < E <  4.82298427 $   \\
 -2.0 & (    5.3313715, 0 ) & $5.33135537   < E <  5.33140311 $  \\
 -1.5 & (    5.8576148, 0 ) & $5.85759578   < E <  5.85766576 $      \\
 -1.0 & (    6.4036752, 0 ) & $6.40362310   < E <  6.40372894 $ \\
  -.5 & (    6.9714613, 0 ) & $6.97133951   < E <  6.97152763 $ \\
   .0 & (    7.5622889, 0 ) & $7.56215901   < E <  7.56242355 $   \\
	&                   & $7.5622738549^3 < E < 7.5622738551^3$ \\
   .5 & (    8.1770143, 0 ) & $8.17687201   < E <  8.17720644 $ \\
  1.0 & (    8.8162241, 0 ) & $8.81569361   < E <  8.81644375 $ \\
  1.5 & (    9.4802011, 0 ) & $9.47913594   < E <  9.48044230 $  \\
  2.0 & (   10.1686764, 0 ) &$10.16683597   < E < 10.16907546 $  \\
  2.5 & (   10.8807993, 0 ) &$10.87896180   < E < 10.88253390 $  \\
  3.0 & (   11.6159802, 0 ) &$11.61535000   < E < 11.62075000 $ \\
  3.5 & (   12.3755952, 0 ) &$12.37090000^4 < E < 12.38980000^4 $ \\
  4.0 & (   13.1638132, 0 ) &$13.16160000^4   < E < 13.20480000^4  $\\
  4.5 & (   13.9842043, 0 ) &$13.86000000^4   < E < 14.04000000^4  $  \\
  5.0 & (   14.8298186, 0 ) &$14.72000000^4   < E < 15.08000000^4  $
\end{tabular}
\noindent{$^1$EMM Analysis of Handy (2001b) \\
$^2$EMM ($P_{max}^{(S)} = 30$) Analysis of Handy (2001a) \\
$^3$EMM ($P_{max}^{(S)} = 50$)  Analysis of Handy (2001a) \\
$^4$EMM ($P_{max}^{(S)} < 30$) Analysis of Handy (2001a)}
\end{center}
\end{table}

\begin{table}
\caption {$E_4$-Branch for the Discrete States of $P^2 + iX^3 + i\alpha X$}
\begin{center}
\begin{tabular}{cccccl}
\multicolumn{1}{c}{$\alpha$}& \multicolumn{1}{c}{$E_4^{(MRF)}(\alpha)$} & \multicolumn{1}{c}{$EMM Bounds: E_R^{(L)} < E_R  < E_R^{(U)}$} \& {$E_I^{(L)} < E_I< E_I^{(U)}$}\\ \hline

 -7.5 & (    4.4103093,    $\pm$ 3.2664292) & $4.408662^1 < E_R < 4.411426^1$,
$3.2645 < E_I < 3.2700$\\
 -7.0 & (    4.3120349,    $\pm$ 2.5634751) & $4.310530 < E_R < 4.313107$,
$2.5618 < E_I < 2.5674$           \\
 -6.5 & (    4.2056380,    $\pm$ 1.9039658) &  $4.204520 < E_R < 4.206360$,
$1.9026 < E_I < 1.9074$             \\
 -6.0 & (    4.1343003,    $\pm$ 1.2850914) & $4.133864 < E_R < 4.134534$, $1.28
4 < \pm E_I < 1.286$         \\
 -5.5 & (    4.1575135,    $\pm$ 0.5313291) & $4.155648 < E_R < 4.158235$,
$.5300 < E_I < .5370$             \\
 -5.0 & (    5.1678291, 0 ) & $5.16784214^2 < E <   5.16798149^2 $      \\
 -4.5 & (    5.8041737, 0 ) & $5.80427900   < E <   5.80439660  $ \\
 -4.0 & (    6.3796826, 0 ) & $6.37969997   < E <   6.37985480 $     \\
 -3.5 & (    6.9490904, 0 ) & $6.94880880   < E <   6.94915872      $ \\
 -3.0 & (    7.5252398, 0 ) & $7.52489539   < E <   7.52536568 $\\
 -2.5 & (    8.1130836, 0 ) & $8.11302413   < E <   8.11368768 $      \\
 -2.0 & (    8.7159047, 0 ) & $8.71631657   < E <   8.71698842   $ \\
 -1.5 & (    9.3364444, 0 ) & $9.33658547   < E <   9.33774669   $ \\
 -1.0 & (    9.9763275, 0 ) & $9.97552896   < E <   9.97718784 $ \\
  -.5 & (   10.6352870, 0 ) &$10.63343056   < E <  10.63660576 $ \\
   .0 & (   11.3120046, 0 ) &$11.31165120   < E <  11.31588480 $ \\
	&	        &$11.314421818^3 < E < 11.314421824^3$ \\		
   .5 & (   12.0072541, 0 ) &$12.01004000   < E <  12.01676000 $ \\
  1.0 & (   12.7264921, 0 ) &$12.72473600   < E <  12.74288000 $ \\
  1.5 & (   13.4761039, 0 ) &$13.47000000   < E <  13.51500000 $ \\
  2.0 & (   14.2487931, 0 ) &$14.21000000   < E <  14.33000000 $  \\
  2.5 & (   15.0088412, 0 ) &$14.96800000   < E <  15.22000000 $ \\
  3.0 & (   15.8347839, 0 ) &$15.71250000   < E <  16.12500000 $ \\
  3.5 & (   16.4020563, 0 ) &$16.40000000   < E <  17.00000000 $  
\end{tabular}
\end{center}
\noindent{$^1$EMM Analysis of Handy (2001b) \\
$^2$EMM ($P_{max}^{(S)} = 30$) Analysis of Handy (2001a) \\
$^3$EMM ($P_{max}^{(S)} = 50$)  Analysis of Handy (2001a) }
\end{table}

\vfil\break

\begin{table}
\caption {MRF Complex Eigenenergy Estimates for $P_{max} = 100$}
\begin{center}
\begin{tabular}{cccccl}
\multicolumn{1}{c}{$\alpha$}& \multicolumn{1}{c}{$E_1 \&\ E_2$} & 
\multicolumn{1}{c}{$E_3 \&\ E_4$} \\ \hline
-7.5 &  & (4.4102443440, $\pm$ 3.2664976653)   \\
-7.0 &  & (4.3119949171, $\pm$ 2.5635584867)   \\
-6.5 &  & (4.2055983584, $\pm$ 1.9040248046)  \\
-6.0 &  & (4.1342519473, $\pm$ 1.2851227083)  \\
-5.5 &  & (4.1574750374, $\pm$ 0.5313567960)  \\
-5.0 &    (1.3433431987, $\pm$ 2.9073906160)  &  \\
-4.5 &    (1.2992423296,$\pm$  2.3125154783) & \\
-4.0 &    (1.2486567335,$\pm$  1.7617193016) & \\
-3.5 &    (1.2124367296,$\pm$  1.2609099725) & \\
-3.0 &    (1.2258475767,$\pm$  0.7600224714) & \\
\end{tabular}
\end{center}
\end{table}

\vfil\break

\begin{table}
\caption {MRF Real Eigenenergy Estimates for $P_{max} = 100$}
\begin{center}
\begin{tabular}{cccccl}
\multicolumn{1}{c}{$\alpha$}& \multicolumn{1}{c}{$E_1$ } &
\multicolumn{1}{c}{$E_2$} & \multicolumn{1}{c}{$E_3$} &
\multicolumn{1}{c}{$E_4$}\\ \hline
-5.0 &      & &
3.4313832016 & 5.1678886857 \\
-4.5 &      & &
3.3232426191 & 5.8043392228 \\
-4.0 &      & &
3.5087656073 & 6.3798052063 \\
-3.5 &      & &
3.8776930052 & 6.9490523010 \\
-3.0 &      & &
4.3334398364 & 7.5251919556 \\
-2.5 & .928000342158 & 1.685979342731 & 4.822975394502 & 8.113377225 \\
-2.0 & .6209135740648 & 2.2922925019 & 5.3313834010 & 8.716619963 \\
-1.5 & .59649338409590 & 2.7425293939422 & 5.8576220019 & 9.337028976 \\
-1.0 & .6999599207986 & 3.1797157763716 & 6.4036451049 &  9.97613238 \\
-0.5 & .8926684335546 & 3.632074461334 & 6.971403910342 & 10.63501626 \\
 0 & 1.1562670719881 & 4.109228752809 & 7.5622738549 & 11.3144218 \\
.5 & 1.47985186079689 & 4.6148387273616 & 8.1770825316 & 12.01482435 \\
1.0 & 1.8561107660566 & 5.150168955614 & 8.8162451717 & 12.73649738 \\
1.5 &  2.2797520475930 & 5.71540870715 & 9.47988920956 &  13.4795632 \\
2.0 & 2.7467399808560 & 6.31023836105 & 10.1679478499 & 14.2440319\\
2.5 & 3.2538769263689 & 6.934096040540 & 10.880225999 & 15.02983115\\
3.0 & 3.798554700716 & 7.586310988692 & 11.616445659 &  15.8368284 \\
3.5 & 4.37859694562365 & 8.2661728690 & 12.37627693 & 16.6648481 \\
4.0 & 4.9921540825786 &  8.9729684347 & 13.15935916 & 17.513684\\
4.5 & 5.637630445616 & 9.7060008600 & 13.96531533 & 18.383109 \\
5.0 & 6.313632040368 & 1.046459957615 & 14.7937618 & 19.2728823\\

\end{tabular}
\end{center}
\end{table}
\vfil\break
\section {Conclusion}
We have confirmed the asymptotic analysis prediction of Delabaerre and Trinh (DT) 
on the existence of symmetry breaking solutions for the $H_\alpha$ 
Hamiltonian. Our methods enable the precise analysis of the complex-real
spectra, particularly for moderate $\alpha$ values at the limits of their (DT)
asymptotic validity. The results of both an eigenenergy estimation method
(MRF) and an eigenenergy bounding method (EMM) were presented. The
algebraic simplicity, and ease of computational implementability, of the
MRF method recommend it highly for application to similar problems. 
Through the use of readily available algebraic programming software, the
MRF approach can be extended to arbitrary precision (indeed, Tymczak et
al (1998b) were able to generate the quartic anharmonic oscillator ground state
energy to more than 250 decimal places), making it a very powerful tool
in these types of investigations.

\vfil\break
\bigskip\centerline {\bf{Acknowledgments}}
This work was supported through a grant from the National Science Foundation
(HRD 9632844) through the Center for Theoretical Studies of Physical Systems
(CTSPS). The author extends his appreciation to Professors D. Bessis,
G. Japaridze, and G. A. Mezincescu for useful discussions.

\vfil\break
\bigskip\centerline {\bf {References}}
\noindent Bender C M and Boettcher S 1998 Phys. Rev. Lett. {\bf 80} 5243

\noindent Bender C M, Berry M, Meisinger P N, Savage V M, and Simsek M 2001
 J. Phys. A: {\bf 34} L31

\noindent Bender C M and Orszag S A,   {\it Advanced Mathematical
Methods for
Scientists and Engineers} (New York: McGraw Hill 1978).

\noindent Delabaere E and Trinh D T 2000 J. Phys. A: Math. Gen. {\bf 33} 8771

\noindent Handy C R and Brooks H 2001 J. Phys. A {\bf }  J. Phys. A {\bf },
to appear.

\noindent Handy C R, Murenzi R, Bouyoucef K, and Brooks H 2000 J. Phys. A 
{\bf 33} 2151

\noindent Handy C R 2001a CAU preprint (Submitted to J. Phys. A)

\noindent Handy C R 2001b CAU preprint (Submitted to J. Phys. A)

\noindent Mezincescu G A 2000 J. Phys. A: Math. Gen. {\bf 33} 4911

\noindent Mezincescu G A 2001 J. Phys. A: Math. Gen. {\bf 34} 3329

\noindent Shohat J A and  Tamarkin J D, {\it The Problem of Moments}
(American Mathematical Society, Providence, RI, 1963).

\noindent Tymczak C J, Japaridze G S, Handy C R, and Wang Xiao-Qian 1998 Phys. Rev. Lett. 80, 3678 

\noindent Tymczak C J, Japaridze G S, Handy C R, and Wang Xiao-Qian 1998 Phys. Rev. A58, 2708

\end{document}